\documentclass[preprint,12pt,authoryear]{elsarticle}

\usepackage{amssymb}
\usepackage{amsmath}
\usepackage{tikz}
\usetikzlibrary{arrows.meta, positioning, shapes.geometric}
\usepackage{hyperref}
\journal{Nuclear Physics B}

\begin{document}

\begin{frontmatter}



\title{Zero-Inflated Model Lab: An Interactive Shiny Application for Modelling Count Data with Excess Zeros} 


\author{Oscar Rodriguez de Rivera Ortega} 

\affiliation{organization={Department of Mathematics and Statistics, University of Exeter},
            country={United Kingdom}}

\begin{abstract}
Count data often contain a large number of zeros in ecological and environmental research, arising from factors such as true species absence, imperfect detection, or the rarity of certain events. Choosing and interpreting suitable statistical models, like Poisson, Negative Binomial, zero‑inflated, or hurdle models, can be difficult for non‑experts because of the models’ assumptions and diagnostic requirements. To address this, we introduce an interactive web application built in R with the Shiny framework that helps users analyse, compare, and interpret count‑data models that may exhibit zero inflation.\\

The application enables users to upload datasets, explore distributions and relationships among variables, and fit a variety of models including Poisson, Negative Binomial, zero‑inflated, and hurdle approaches. It incorporates tools for model comparison using information criteria, visualization of coefficient estimates and confidence intervals, assessment of variable importance, and diagnostic checks based on simulated residuals. Importantly, the app provides clear visualizations of both the count‑generating and zero‑generating processes, making it easier to communicate components that are typically challenging to interpret. \\

To improve accessibility, the tool offers guided explanations of outputs and diagnostics, making it valuable for both research applications and teaching. A case study using environmental count data illustrates how the application helps users identify zero inflation, choose appropriate models, and gain insight into the ecological processes underlying the data.\\

Overall, the application narrows the gap between advanced statistical methods and practical analysis, supporting reproducible workflows and enhancing the clarity and interpretability of models for zero‑inflated count data.

\end{abstract}









\end{frontmatter}



\section{Introduction}

Ecological datasets frequently contain response variables recorded as counts, count-like measurements, or non-negative durations. Examples include species abundance \citep{welsh1996modelling, cunningham2005modeling}, numbers of damaged individuals \citep{madrigal2023empirical, hood2018fire}, regeneration counts \citep{rathbun2006spatial}, infection events \citep{bolker2009generalized}, fire scars \citep{madrigal2023empirical}, mortality events \citep{hood2018fire}, and exposure durations above biologically meaningful thresholds \citep{madrigal2023empirical, espinosa2020predicting}. Such responses often violate the assumptions of standard Gaussian models because they are discrete or semi-discrete, bounded below by zero, right-skewed, and frequently characterised by a high proportion of zero observations \citep{welsh1996modelling, martin2005zero, warton2005many, bolker2009generalized, ohara2010do}. In ecology, zeros are rarely a purely statistical inconvenience. Instead, they may reflect different ecological and observational processes, including true absence, lack of exposure, unsuitable environmental conditions, non-detection, limited sampling effort, or structural constraints that make a positive response impossible \citep{martin2005zero}. Therefore, modelling the source of zero observations is essential for robust ecological inference.

Zero-inflated data are especially common because many ecological processes are naturally sparse in space and time \citep{silva2025joint}. Rare species may be absent from many sampling units, regeneration may occur only where environmental filters allow recruitment, disease or mortality events may only appear under specific biological conditions, and fire damage may only be observed when individuals are exposed to sufficient heat. In such cases, observed zeros may correspond to true ecological zeros, false zeros caused by imperfect detection, random zeros generated by low expected abundance, or structural zeros produced by ecological constraints \citep{cunningham2005modeling, martin2005zero}. Treating all zeros as equivalent can obscure the mechanisms generating the data and may lead to biased parameter estimates, poor predictions, or misleading ecological interpretation \citep{blasco2019zero}.

However, the presence of many zeros does not automatically imply that a zero-inflated model is required. Ecological datasets with low expected counts may naturally contain many zeros even under standard count distributions, and overdispersion can sometimes be addressed using negative binomial models rather than explicitly modelling a separate zero-generating process \citep{warton2005many, zeileis2008regression}. This distinction is important because zero-inflated and hurdle models imply different assumptions about the data-generating process. Zero-inflated models assume that zeros may arise from either a systematic zero-generating component or a count-generating process, whereas hurdle models separate the process into a mechanism that generates zeros and a mechanism that generates non-zero positive responses \citep{cragg1971some, mullahy1986specification, lambert1992zero, heilbron1994zero, zeileis2008regression}. Thus, the choice of model should be guided not only by the proportion of zeros, but also by exploratory analysis, model comparison, residual diagnostics, and ecological reasoning.


A direct motivation for the present work is the study by \citet{madrigal2023empirical}, where we developed an empirical modelling framework for stem cambium heating caused by prescribed burning in Mediterranean pine forests. In that study, the duration of cambium temperature above $60^\circ$C, denoted $tC_{60}$, was used as a proxy for potential cambium damage at the tree scale. The authors showed that tree traits, fire severity indicators, and burn season were associated with both the probability and duration of cambial heating. In particular, bark thickness acted as a protective trait, while variables such as crown scorch, stem scorch, and maximum bark temperature were associated with increased thermal exposure. Importantly, the study used a hurdle modelling framework, highlighting the need to distinguish between trees with no damaging thermal exposure and trees with positive exposure durations. This modelling problem provided a key motivation for developing an accessible online platform that allows users to explore, fit, compare, and interpret alternative count-data models for ecological responses with excess zeros.

From a statistical perspective, Poisson models provide a useful baseline for count responses, but they assume equality between the conditional mean and variance. This assumption is frequently violated in ecological data, where overdispersion is common due to unobserved heterogeneity, clustering, environmental variability, or missing covariates \citep{bolker2009generalized, ohara2010do, zeileis2008regression}. Negative binomial models relax this assumption by introducing an additional dispersion parameter. However, when overdispersion is driven partly by an excess of zero observations, zero-inflated or hurdle formulations may provide a more ecologically meaningful representation of the underlying process \citep{lambert1992zero, mullahy1986specification, sileshi2009traditional}. These models have been applied in a range of ecological and environmental contexts, including rare species abundance, spatial count data, oak regeneration, and conservation problems involving rare events \citep{welsh1996modelling, agarwal2002zero, rathbun2006spatial, cantoni2017random}.

Despite the availability of these methods, selecting and interpreting an appropriate model remains challenging in applied ecological research. Researchers must inspect the distribution of the response variable, evaluate the prevalence and possible origin of zeros, assess overdispersion, explore relationships among predictors, compare competing models, and check residual diagnostics. Model comparison using information criteria, such as the Akaike Information Criterion (AIC), provides a useful basis for comparing candidate models, but it should not be used in isolation \citep{akaike1974new}. A model with lower AIC may still be ecologically implausible, poorly calibrated, or affected by unmodelled structure. Therefore, model selection should be embedded within a broader workflow that includes graphical exploration, diagnostic checks, uncertainty assessment, prediction evaluation, and ecological interpretation.

Interactive software can help address this need by making complex modelling workflows more transparent, reproducible, and accessible. Web applications developed in \texttt{R} using \texttt{Shiny} allow users to interact with statistical models through a graphical interface while retaining the flexibility of established statistical packages \citep{chang2024shiny}. This is particularly valuable in ecological informatics, where researchers, students, and practitioners often need to analyse complex ecological datasets but may not always have advanced programming expertise. An interactive application can guide users through the complete modelling process, from data upload and exploratory analysis to model fitting, comparison, diagnostics, prediction, and interpretation.

In this paper, we present an interactive \texttt{R Shiny} application for the analysis of ecological count data with potential overdispersion and excess zeros. The application allows users to upload a dataset, inspect its structure, explore the response distribution and predictor relationships, fit alternative count models, compare models using AIC, and interpret fitted models through coefficient plots, variable importance measures, partial dependence plots, residual diagnostics, predicted-versus-observed plots, and zero-probability visualisations. The modelling framework includes Poisson, negative binomial, zero-inflated negative binomial, and hurdle models, implemented using standard \texttt{R} tools including \texttt{glm}, \texttt{MASS}, and \texttt{glmmTMB} \citep{venables2002modern, brooks2017glmmtmb}. Model diagnostics are supported by simulation-based residual checks using \texttt{DHARMa}, while graphical outputs are generated using \texttt{ggplot2} and related visualisation tools \citep{hartig2024dharma, wickham2016ggplot2}.

The application is illustrated using a simulated post-fire tree damage dataset inspired by the modelling problem described by \citet{madrigal2023empirical}. In the example, the response variable represents the duration for which cambium temperature exceeded $60^\circ$C. Some trees do not experience heating above this threshold, generating zero observations, whereas positive values vary according to fire exposure indicators and tree resistance traits. Although the case study focuses on post-fire tree damage, the workflow is broadly applicable to many ecological count and count-like responses, including abundance data, damage scores, event counts, infection incidence, mortality counts, and exposure-duration outcomes. The main contribution of the application is therefore to provide an accessible, reproducible, and interpretable modelling workflow that helps users connect statistical model choice with ecological process understanding.

\section{Application design}

\subsection{Application structure}

\begin{figure}[htbp]
\centering
\begin{tikzpicture}[
    node distance=1.4cm,
    every node/.style={font=\small},
    box/.style={rectangle, draw, rounded corners, minimum width=3.4cm, minimum height=1cm, align=center},
    arrow/.style={->, thick}
]

\node[box] (input) {User input\\Upload data and select variables};
\node[box, below=of input] (explore) {Data exploration\\Histograms, factor plots, pairwise plots};
\node[box, below=of explore] (model) {Model fitting\\Poisson, NB, ZINB, hurdle};

\node[box, below left=1.8cm and 2.2cm of model] (compare) {Model comparison\\AIC};
\node[box, below=1.8cm of model] (outputs) {Model outputs\\Summaries, coefficients, importance};
\node[box, below right=1.8cm and 2.2cm of model] (predict) {Predictions\\Observed vs predicted};

\node[box, below=1.8cm of outputs] (diagnostics) {Diagnostics\\DHARMa residuals and zero diagnostics};

\draw[arrow] (input) -- (explore);
\draw[arrow] (explore) -- (model);
\draw[arrow] (model) -- (compare);
\draw[arrow] (model) -- (outputs);
\draw[arrow] (model) -- (predict);
\draw[arrow] (outputs) -- (diagnostics);
\draw[arrow] (predict) -- (diagnostics);
\draw[arrow] (compare) -- (diagnostics);

\end{tikzpicture}
\caption{General scheme of the application structure.}
\label{fig1}
\end{figure}
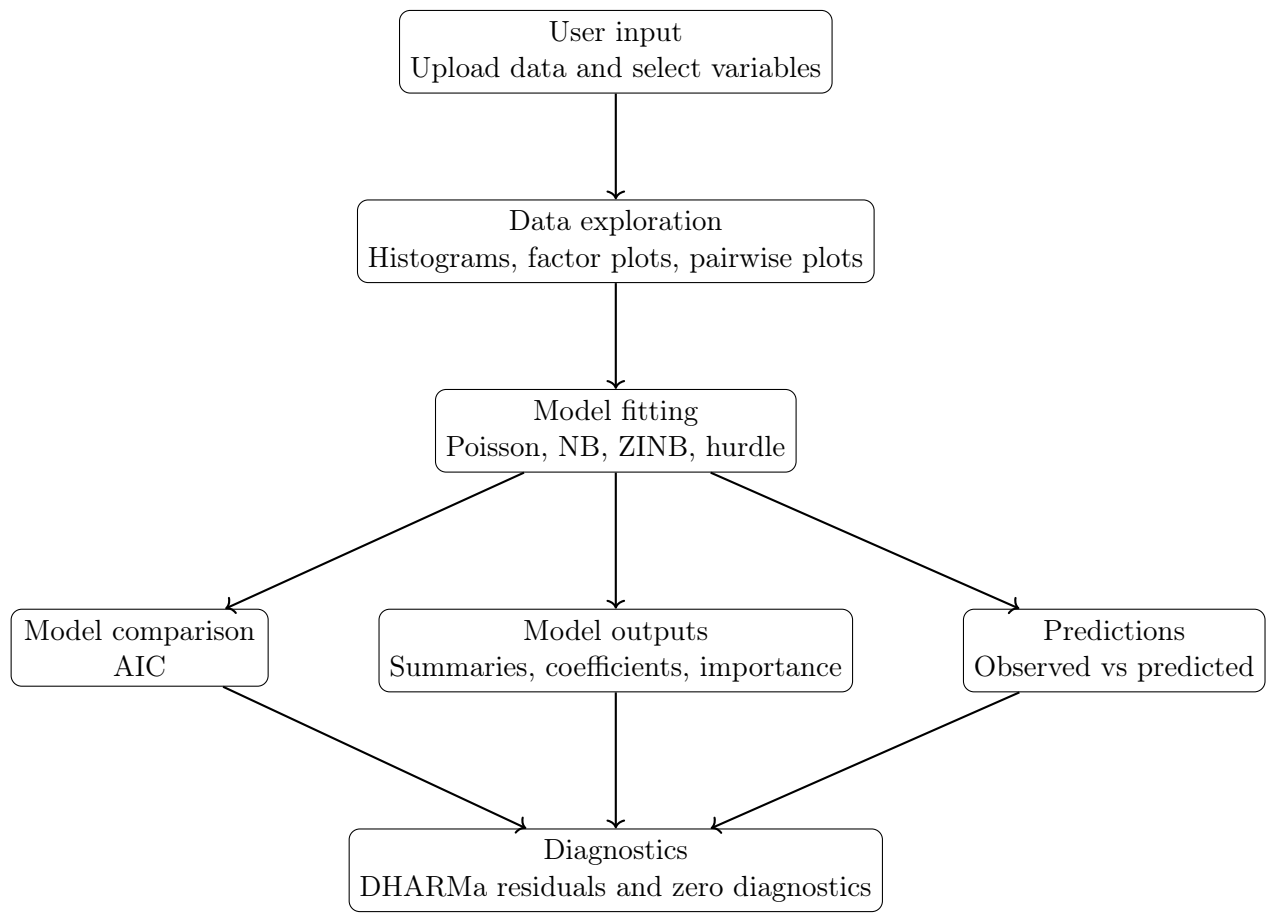

The design of \textit{Zero-Inflated Model Lab} focuses on enabling users to analyse complex ecological count data without requiring advanced programming expertise, while also providing guidance on model selection, interpretation, and diagnostic assessment. The application integrates the main steps of a count-data modelling workflow, from data upload and exploratory analysis to model fitting, comparison, prediction, and diagnostics. A schematic overview of the application structure is shown in Fig.~\ref{fig1}. The interface is organised into the following tabs, each corresponding to a specific stage of the analytical workflow.

\begin{enumerate}

\item \textbf{Data}. This tab provides an initial overview of the uploaded dataset. Users can inspect both the structure and summary statistics of the data, including variable types, ranges, and basic descriptive measures. This step helps verify that the dataset has been uploaded correctly and that the variables are appropriately formatted for count-based modelling.

\item \textbf{Exploration}. This tab provides exploratory visualisations to assess the data before model fitting. It includes a histogram of the selected response variable, which allows users to examine its distribution, skewness, dispersion, and the presence of excess zeros. It also includes pairwise plots of numerical predictors, helping users identify potential relationships, correlations, outliers, and patterns that may influence model specification.

\item \textbf{Factor plots}. This tab focuses on the relationship between categorical predictors and the response variable. It allows users to compare the response across factor levels using boxplots and individual observations. These visualisations can help identify group-level differences, heterogeneity among categories, and potential categorical effects that may be relevant for inclusion in the model.

\item \textbf{Data diagnostics}. This tab provides basic diagnostic summaries of the uploaded dataset and the selected response variable. These outputs help users identify potential data-quality issues, such as missing values, unusual values, or distributional features that may affect model fitting and interpretation. The tab also provides preliminary guidance on the type of model that may be suitable for the observed data structure.

\item \textbf{Zero diagnostics}. This tab is specifically designed to assess the prevalence of zero values in the response variable. It reports the proportion of zero observations and displays a graphical summary of zero and non-zero responses. This information helps users evaluate whether standard count models may be sufficient, or whether more flexible approaches, such as zero-inflated or hurdle models, should be considered.

\item \textbf{Model comparison}. This tab allows users to compare a set of candidate models, including Poisson, negative binomial, zero-inflated negative binomial, and hurdle models. Model performance is evaluated using the Akaike Information Criterion (AIC), which provides a measure of the trade-off between model fit and complexity. The model with the lowest AIC is highlighted as the best-supported model according to this criterion.

\item \textbf{Model summary}. This tab presents the numerical output for the selected model, including parameter estimates, standard errors, test statistics, and significance values. For zero-inflated and hurdle models, results are displayed separately for the count and zero components, allowing users to interpret both the intensity of the response and the process associated with zero outcomes.

\item \textbf{Variable importance}. This tab provides a graphical summary of the relative contribution of each predictor to the selected model. Importance is calculated using the absolute magnitude of the estimated coefficients. This allows users to identify predictors with stronger estimated effects, although interpretation should account for predictor scaling and uncertainty.

\item \textbf{Coefficient plot}. This tab displays model coefficients together with approximate confidence intervals. This visualisation helps users assess the direction, magnitude, and uncertainty of predictor effects. For zero-inflated and hurdle models, coefficient plots are separated by model component, making it easier to compare effects in the count and zero processes.

\item \textbf{Partial dependence}. This tab visualises the marginal effect of an individual predictor on the predicted response. For numerical predictors, predictions are generated across the observed range of the selected variable while holding other variables constant. For categorical predictors, predicted values are compared across factor levels. These plots help users interpret model behaviour on the response scale.

\item \textbf{Residual diagnostics}. This tab provides simulation-based residual diagnostics using the \texttt{DHARMa} framework. These diagnostics help assess whether the fitted model adequately captures the structure of the data and whether there is evidence of model misspecification, overdispersion, or systematic residual patterns.

\item \textbf{Prediction}. This tab compares observed and predicted values for the selected model. The resulting plot allows users to assess how well the model reproduces the observed response and to identify systematic overprediction, underprediction, or poor fit for particular regions of the response distribution.

\item \textbf{Zero probability}. This tab is available for zero-inflated and hurdle models and focuses on the zero-generating component of the model. It displays the predicted probability of a zero outcome for each observation, providing insight into the conditions under which zero responses are more likely to occur.

\end{enumerate}

Together, these components support an iterative modelling workflow. Users can explore the data, fit alternative model structures, compare model performance, inspect parameter estimates, evaluate predictions, and assess diagnostics before refining the model specification. This is particularly useful for ecological datasets characterised by overdispersion, excess zeros, and heterogeneous predictor effects.

For more advanced users, the application also promotes transparency and reproducibility by exposing the main components of the modelling workflow, including the selected response variable, predictor set, zero-part predictors, model formulae, and fitted model outputs. As a result, analyses conducted within the application can be reproduced, adapted, or extended outside the app environment using standard \texttt{R} tools.

\section{Presentation and GitHub repository}

\begin{figure}[htbp]
    \centering
    \includegraphics[width=1\textwidth]{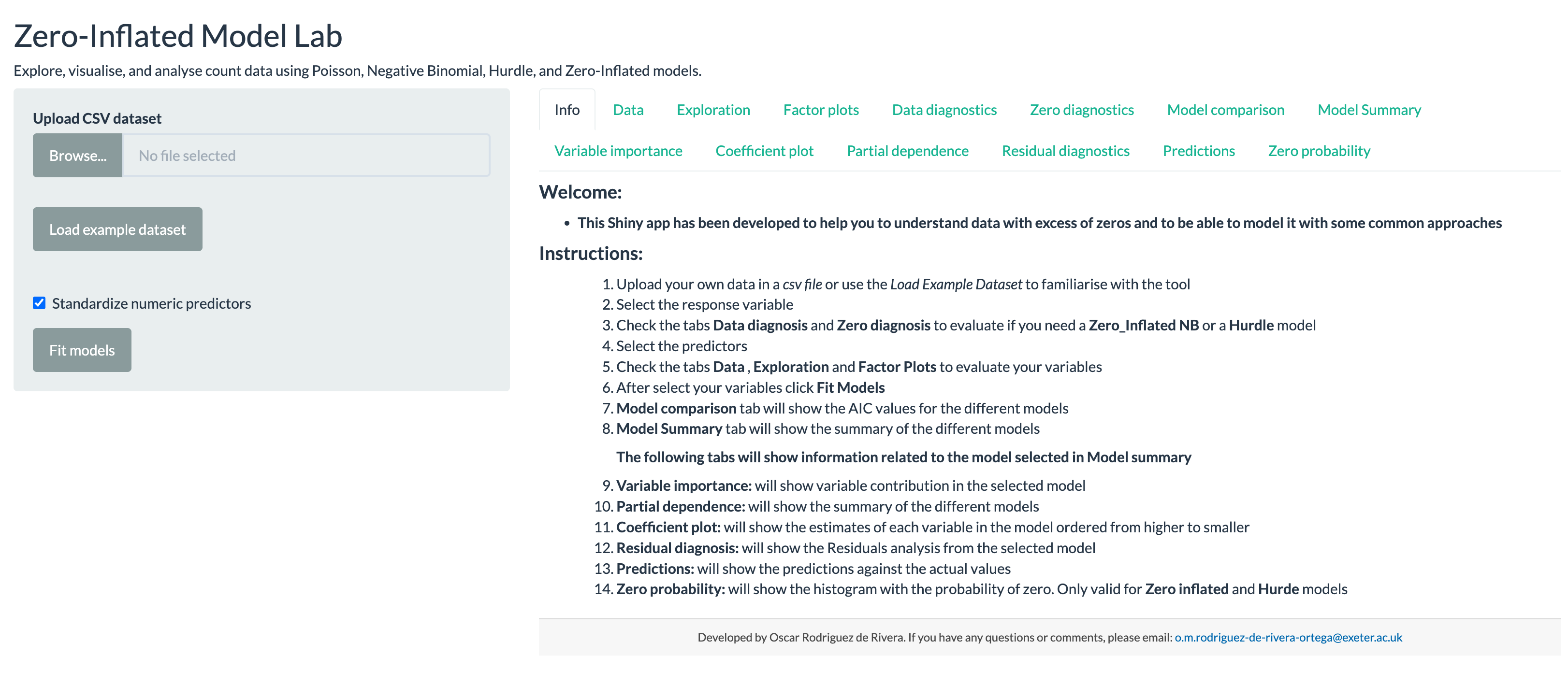}
    \caption{Detail of the app Info window.}
    \label{fig:infoapp}
\end{figure}

As can be seen in Fig.~\ref{fig:infoapp}, the info tab briefly highlights the motivation, context for the development of the app, which is a summary of the abstract of this article. Also, it explains step by step the use of the app, from using the data example/upload the data to obtain predictions of zeros.

The app is available through the following GitHub repository: \url{https://github.com/orrortega/zero-inflated-model-lab}, and can also be accessed directly as a web app via

\url{https://orderortega.shinyapps.io/zero-inflated-model-lab/}. 

The repository provides useful information for users, including guidance on how to use the app and explanations that support a deeper understanding of the modelling structures implemented within it.

Through the GitHub page, users can request changes, suggest new implementations, or report bugs.

\section{Simulated data}

A synthetic tree-level dataset was generated to illustrate the proposed modelling framework. The dataset comprised $n = 400$ observations collected across 20 sites, with 20 trees sampled per site. Each observation represented an individual tree and included information on species identity, tree structure, fire exposure, and burning conditions. Species identity was represented by three pine species: \textit{Pinus nigra}, \textit{Pinus pinaster}, and \textit{Pinus pinea}. Structural covariates included diameter at breast height (DBH, cm), bark thickness (cm), and total tree height (m). Fire-related covariates included crown scorch percentage, maximum stem char height (m), maximum bark temperature, and burn season.

The response variable, $tC60_i$, represented the duration for which the cambium temperature of tree $i$ exceeded $60^\circ$C. This variable was generated using a two-stage process designed to reproduce a common feature of post-fire ecological data: a large number of zero observations combined with positive count values for affected trees. The first stage determined whether a tree experienced a non-zero cambial heating event, while the second stage generated the duration of heating conditional on such an event occurring.

Let $CS_i$ denote crown scorch percentage, $BT_i$ bark thickness, $BTM_i$ maximum bark temperature, and $SCH_i$ maximum stem char height for tree $i$. We define $z_i$ as a binary indicator taking value 1 if tree $i$ experienced cambial heating above the threshold and 0 otherwise. The probability of a non-zero fire-impact event was generated as

\begin{equation}
P(z_i = 1) =
\mathrm{logit}^{-1}
\left(
-2 + 0.03CS_i + 0.002BTM_i - 0.4BT_i
\right).
\end{equation}

This formulation assumes that the probability of cambial heating increases with crown scorch percentage and maximum bark temperature, reflecting greater fire exposure, and decreases with bark thickness, reflecting the insulating role of bark.

Conditional on a non-zero fire-impact event, the duration of cambial heating above $60^\circ$C was generated from a Poisson distribution,

\begin{equation}
tC60_i \mid z_i = 1 \sim \mathrm{Poisson}(\lambda_i),
\end{equation}

where

\begin{equation}
\log(\lambda_i) =
0.2 + 0.04CS_i + 0.3SCH_i - 0.2BT_i.
\end{equation}

Thus, among trees experiencing cambial heating, the expected duration increased with crown scorch percentage and stem char height, and decreased with bark thickness. Trees that did not experience a fire-impact event were assigned a zero response,

\begin{equation}
tC60_i = 0 \quad \mathrm{if} \quad z_i = 0.
\end{equation}

The resulting response therefore follows a hurdle-type data-generating process, in which the occurrence of cambial heating and the duration of heating conditional on occurrence are treated as distinct ecological processes. In this example, crown scorch, stem char height, and maximum bark temperature act as indicators of fire exposure or severity, whereas bark thickness represents a tree-level resistance trait that reduces both the probability and expected duration of damaging cambial heating.

\section{Uploading the data}

\begin{figure}[htbp]
    \centering
    \includegraphics[width=1\textwidth]{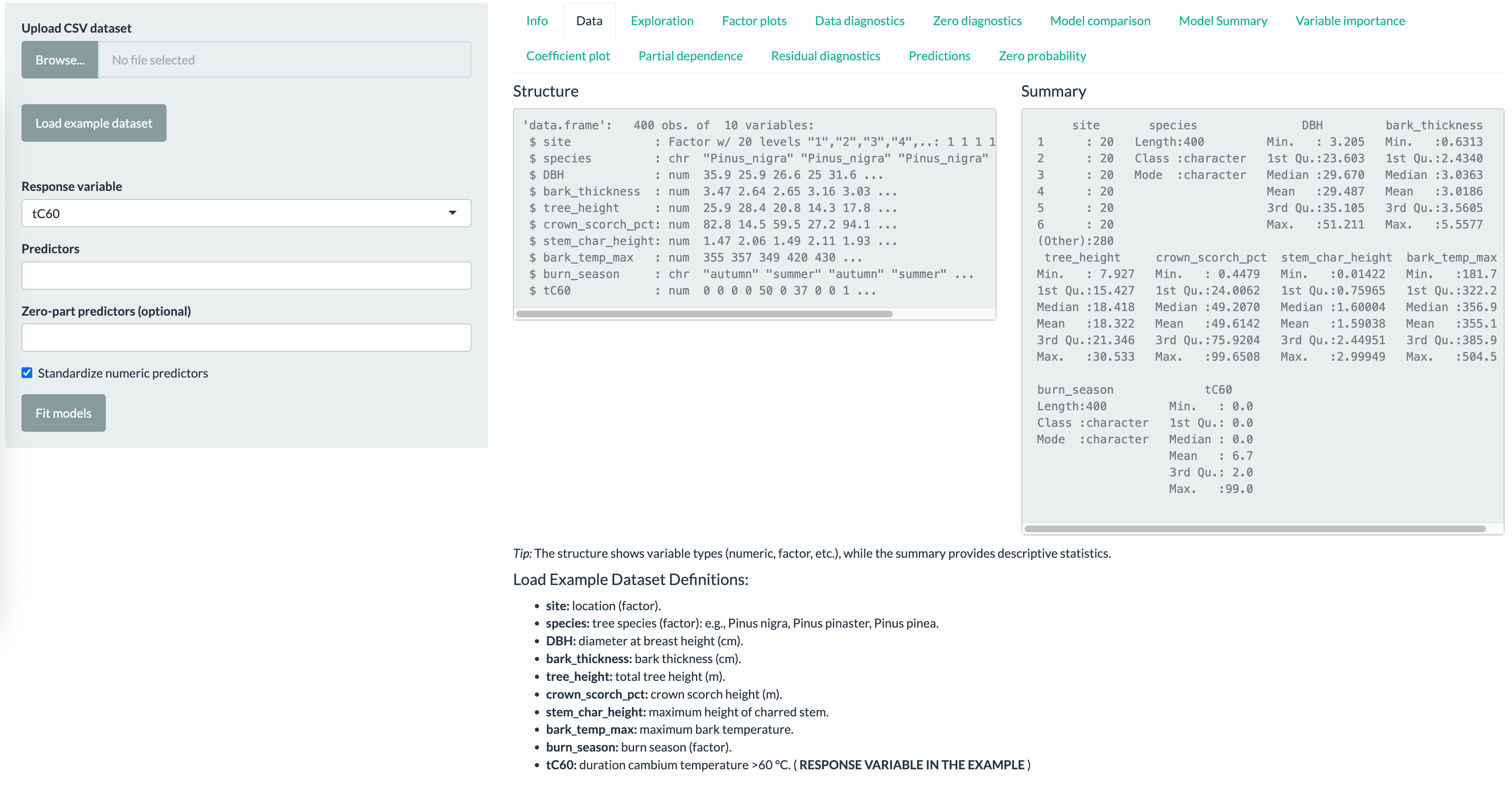}
    \caption{The Uploading data tab allows data be imported for analysis, including the response variable and explanatory variables.}
    \label{fig:dataapp}
\end{figure}

This section describes the data upload component of the application, including the structure of the user interface and the way in which uploaded data are presented through the \textit{Data} tab (Fig.~\ref{fig:dataapp}).

The first input element of the interface allows users to upload the dataset required for the analysis. Data must be provided in comma-separated values format (\texttt{.csv}) to ensure that the file is read correctly by the application.

After the dataset has been uploaded, the \textit{Data} tab displays two diagnostic summaries. The first provides the structure of the dataset using the \texttt{str()} output, including the number of observations, the number of variables, and the data type of each variable, such as character, factor, or numeric. The second provides a summary of the variables using the \texttt{summary()} output. For numerical variables, this includes the minimum, first quartile, median, mean, third quartile, and maximum values. For categorical variables, the summary reports the distribution of observations across the available categories (\textbf{reference for figure}).

Together, these outputs allow users to verify that the dataset has been uploaded correctly and to inspect the main characteristics of the variables before proceeding with the exploratory and modelling stages of the analysis.

\section{Data exploration and descriptive analysis}

\subsection{Data exploration and descriptive analysis}

\begin{figure}[htbp]
    \centering
    \includegraphics[width=1\textwidth]{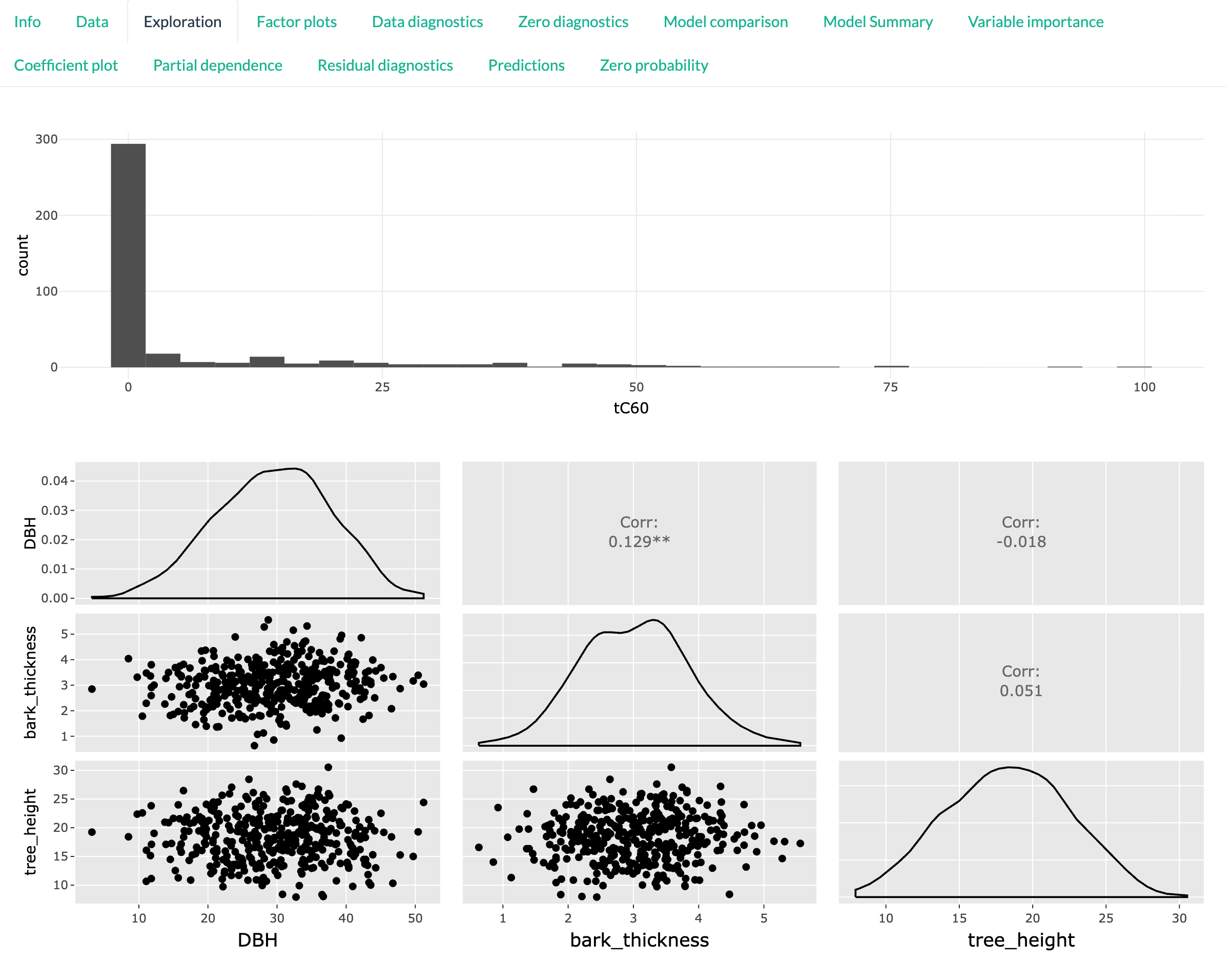}
    \caption{The Exploration tab allows the user to see the distribution of the selected response variable and a scatterplot matrix of continuous predictors.}
    \label{fig:explorationapp}
\end{figure}

The \textit{Exploration} tab of the application provides a set of descriptive and diagnostic tools to support the preliminary assessment of the uploaded dataset before model fitting. These outputs are designed to help users explore the distribution of the response variable, inspect relationships among predictors (categorical and continuous), and identify data characteristics that may inform the choice of an appropriate count model (Fig.~\ref{fig:explorationapp}).

The first output is an interactive histogram of the selected response variable. The histogram is generated using 30 bins and displayed through an interactive \texttt{plotly} interface. This visualisation allows users to examine the empirical distribution of the response, including skewness, dispersion, and the presence of a high frequency of zero values. For ecological count responses, such as the duration of cambial heating above a threshold, a right-skewed distribution is often expected. However, a large concentration of observations at zero may indicate that a standard Poisson or negative binomial model may not fully capture the data-generating process, motivating the consideration of zero-inflated or hurdle models.

The second output is an interactive scatterplot matrix of the selected numerical predictors. This plot is produced using the \texttt{ggpairs()} function from the \texttt{GGally} package and converted to an interactive display using \texttt{plotly}. Only numerical variables that have also been selected as predictors are included in the matrix. The diagonal panels summarise the marginal distribution of each predictor, while the off-diagonal panels display pairwise relationships between predictors. This visualisation helps users detect strong correlations, non-linear patterns, skewed predictor distributions, and potential outliers. Such features are important because collinearity or strong non-linear relationships can affect model stability and the interpretation of covariate effects.

\begin{figure}[htbp]
    \centering
    \includegraphics[width=1\textwidth]{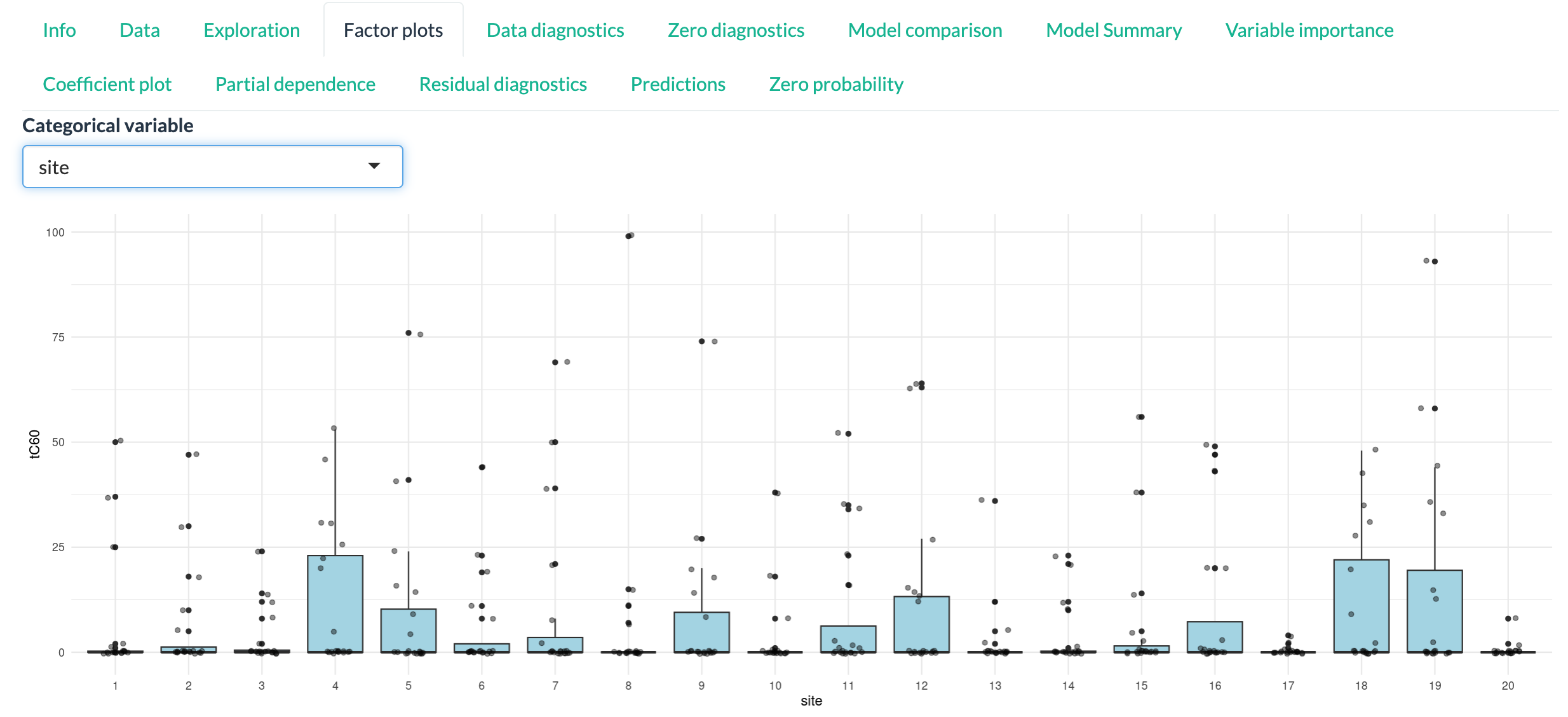}
    \caption{The Factor plots tab allows the user to select and understand graphically the different categorical variables.}
    \label{fig:factorapp}
\end{figure}

A separate \textit{Factor plots} tab is provided for categorical predictors. The application automatically identifies variables stored as factors or character strings and allows the user to select one categorical variable at a time. For the selected variable, the response is displayed using boxplots with overlaid jittered observations. This representation allows users to compare the distribution of the response across categories, such as species or burn season, while also showing the underlying individual observations. These plots are useful for identifying differences between groups, unequal variability among factor levels, and categories with sparse observations (Fig.~\ref{fig:factorapp}).

.

\begin{figure}[htbp]
    \centering
    \includegraphics[width=1\textwidth]{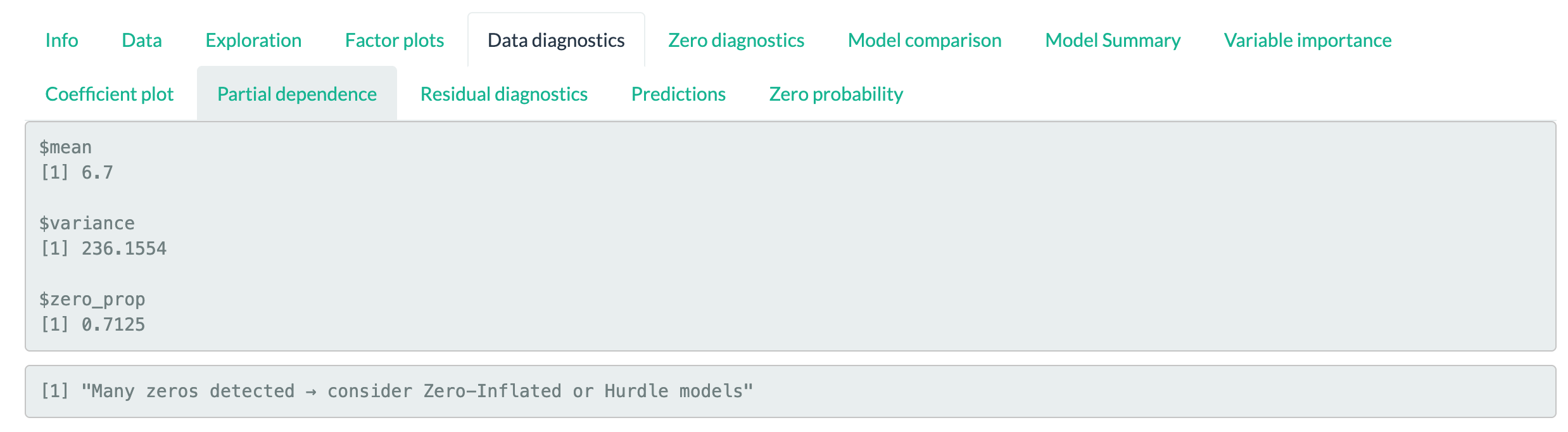}
    \caption{The Data diagnostics tab presents and recommends a preliminary model according to the response variable}
    \label{fig:datadiagnosisapp}
\end{figure}

The application also includes a \textit{Data diagnostics} tab, which provides numerical summaries of data quality for the selected response variable. These diagnostics are computed using the \texttt{data\_quality()} function and are then passed to the \texttt{recommend\_model()} function to produce a preliminary model recommendation. Although this recommendation is intended as guidance rather than a substitute for formal model comparison, it helps users identify whether the response shows features such as excess zeros, overdispersion, or other characteristics that may favour Poisson, negative binomial, zero-inflated, or hurdle-type models (Fig.~\ref{fig:datadiagnosisapp}).

.

\begin{figure}[htbp]
    \centering
    \includegraphics[width=1\textwidth]{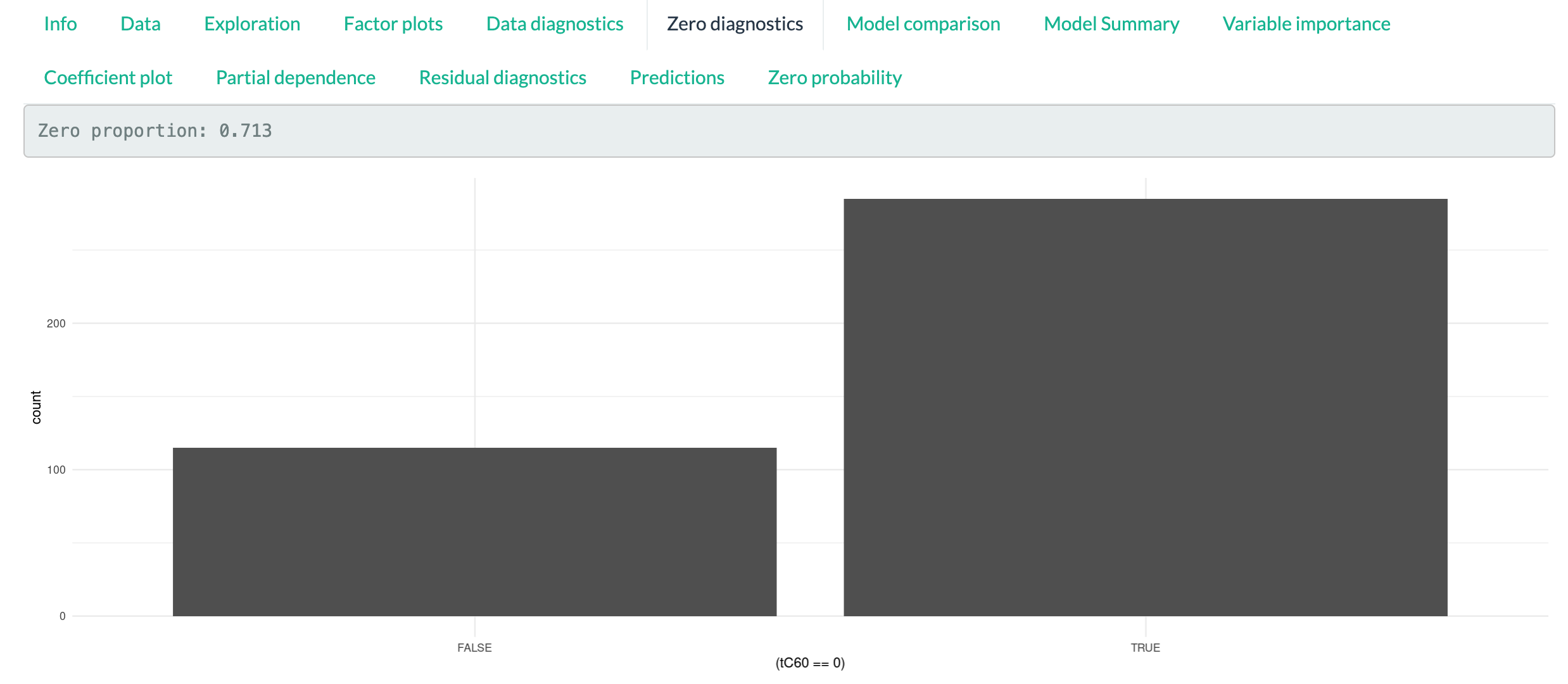}
    \caption{The Zero diagnosis reports the proportion of observations equal to zero and displays a bar plot separating zero and non-zero responses}
    \label{fig:zerodiagnosisapp}
\end{figure}

Finally, the \textit{Zero diagnostics} tab provides an assessment of the prevalence of zero values in the response variable. The application reports the proportion of observations equal to zero and displays a bar plot separating zero and non-zero responses. This diagnostic is particularly relevant for ecological count data, where zeros may arise from different ecological processes, such as true absence, lack of exposure, non-detection, or structural constraints. A high proportion of zeros may therefore suggest the need for a modelling framework that explicitly distinguishes between the occurrence of the process and the intensity of the response conditional on occurrence (Fig.~\ref{fig:zerodiagnosisapp}).

.

Overall, the exploratory analysis module provides an initial assessment of the response distribution, predictor structure, categorical effects, and zero prevalence. These outputs support a more informed modelling workflow by allowing users to identify key data features before selecting and fitting candidate statistical models.

\subsection{Variable selection, model fitting and comparison}

Following the exploratory analysis, the application allows users to define, fit and compare alternative count models for the selected response variable. The model fitting workflow is controlled through a set of interactive input elements that are generated automatically once a dataset has been uploaded.

First, the user selects the response variable, denoted by $Y_i$, from the list of available variables in the dataset. This variable represents the ecological outcome of interest, such as the duration for which cambium temperature exceeded $60^\circ$C. Once the response variable has been selected, the remaining variables are made available as candidate predictors for the count-generating component of the model. Users can then select one or more explanatory variables, denoted by $\mathbf{x}_i$, using a multiple-selection input.

\begin{figure}[htbp]
    \centering
    \includegraphics[width=0.5\textwidth]{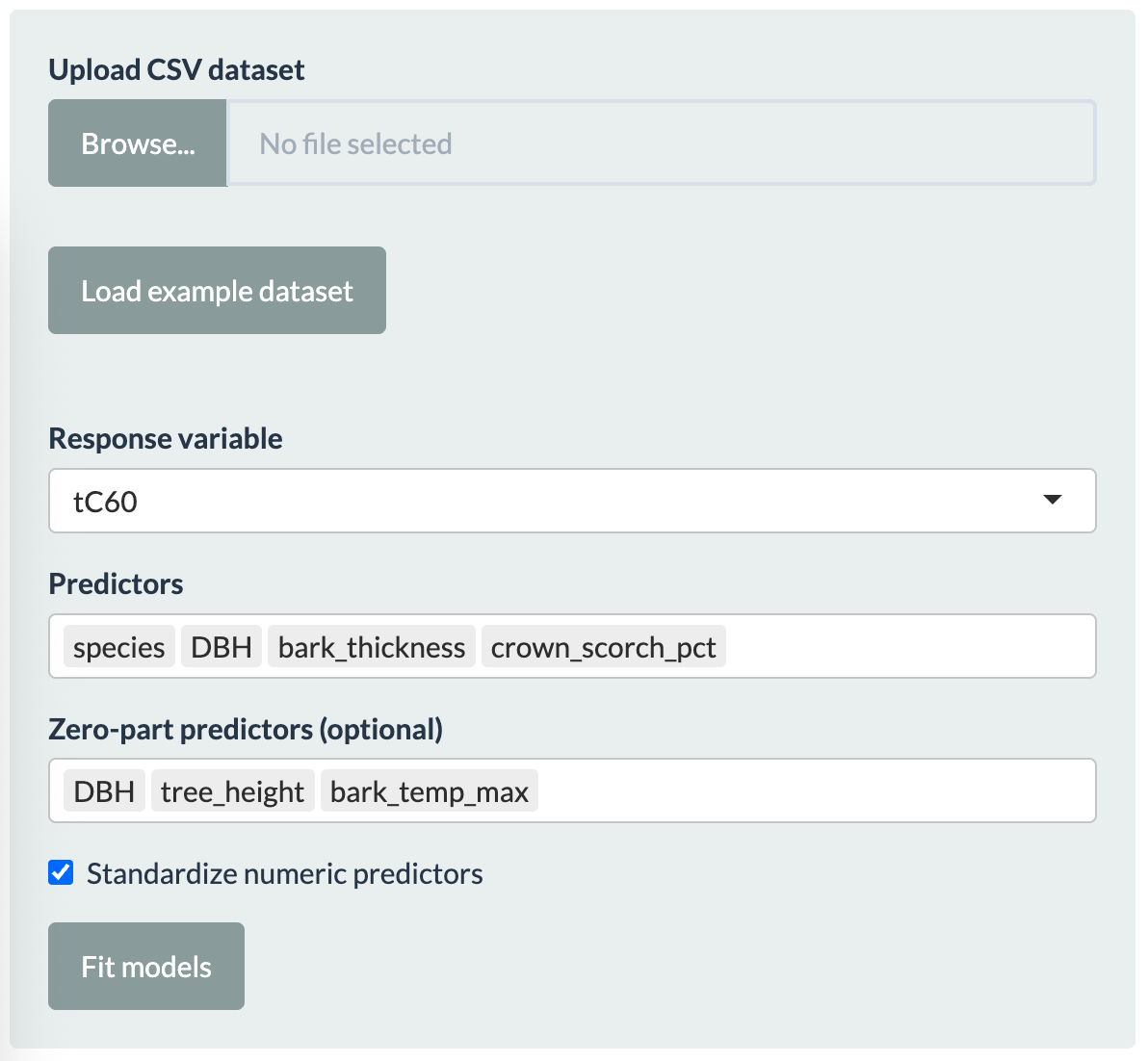}
    \caption{Interface elements for determining which data to analyse (simulated or uploaded) and select which variables upload to the predictors.}
    \label{fig:uploadfitapp}
\end{figure}

For models that include a zero-generating process, the application also provides an optional input for selecting zero-part predictors. These variables, denoted by $\mathbf{w}_i$, are used to model the probability of excess zeros in the zero-inflated negative binomial model, or the probability of crossing the zero hurdle in the hurdle model. This separation allows users to test whether the processes governing the occurrence of zero observations differ from those controlling the magnitude of positive counts (Fig.~\ref{fig:uploadfitapp}).

Before model fitting, the application performs a clean-up by retaining only the selected response and predictor variables and removing observations that cannot be used in the analysis. Users may also choose to standardise numerical predictors (but not the response variable). This option is useful when predictors are measured on different scales, as it allows estimated effects to be compared more directly and can help estimation stability.

The general count component of the fitted models is defined as
\begin{equation}
\log(\mu_i) = \beta_0 + \mathbf{x}_i^\top \boldsymbol{\beta},
\end{equation}
where $\mu_i = E(Y_i)$ is the expected value of the response for observation $i$, $\beta_0$ is the intercept, and $\boldsymbol{\beta}$ is the vector of regression coefficients associated with the selected predictors. The log-link function ensures that predicted mean values are positive, as required for count data.

Four candidate models are fitted automatically: a Poisson model, a negative binomial model, a zero-inflated negative binomial model, and a hurdle model. The Poisson model is fitted as a baseline count model,
\begin{equation}
Y_i \sim \mathrm{Poisson}(\mu_i),
\end{equation}
and assumes equality between the conditional mean and variance. Although this model provides a useful starting point, it may be inappropriate when the data exhibit overdispersion. To account for overdispersion, the application also fits a negative binomial model,
\begin{equation}
Y_i \sim \mathrm{NegBin}(\mu_i, \theta),
\end{equation}
where $\theta$ is an additional dispersion parameter (conventionally called the ``size'' parameter). This model relaxes the Poisson mean--variance assumption and is therefore more suitable when the observed variance exceeds the mean.

The zero-inflated negative binomial model extends the negative binomial formulation by allowing zeros to arise from two sources: a structural-zero process and a count-generating process. The zero component is modelled as
\begin{equation}
\mathrm{logit}(\pi_i) = \gamma_0 + \mathbf{w}_i^\top \boldsymbol{\gamma},
\end{equation}
where $\pi_i$ is the probability that observation $i$ belongs to the structural-zero component, $\mathbf{w}_i$ is the vector of selected zero-part predictors, and $\boldsymbol{\gamma}$ is the corresponding vector of coefficients. This results in a mixture distribution for the counts, where zeros are generated with probability:
\begin{equation}
P(Y_i = 0) = \pi_i + (1 - \pi_i)P_{\mathrm{NB}}(Y_i = 0),
\end{equation}
whereas non-zero counts arise with probability:
\begin{equation}
P(Y_i = y_i) = (1 - \pi_i)P_{\mathrm{NB}}(Y_i = y_i),
\quad y_i > 0.
\end{equation}

Finally, the application fits a hurdle model using a zero-truncated negative binomial distribution for the positive counts. In this case, the zero and positive components are treated as two distinct processes. The zero component describes whether the response crosses the zero ``hurdle'',
\begin{equation}
P(Y_i = 0) = \pi_i,
\end{equation}
whereas the positive component models the response conditional on being greater than zero,
\begin{equation}
P(Y_i = y_i \mid Y_i > 0) =
P_{\mathrm{TNB}}(Y_i = y_i),
\quad y_i > 0.
\end{equation}
Here, $P_{\mathrm{TNB}}$ denotes the zero-truncated negative binomial distribution. This formulation is particularly relevant for ecological count data when zero values and positive counts are thought to arise from different ecological mechanisms.

\begin{figure}[htbp]
    \centering
    \includegraphics[width=1\textwidth]{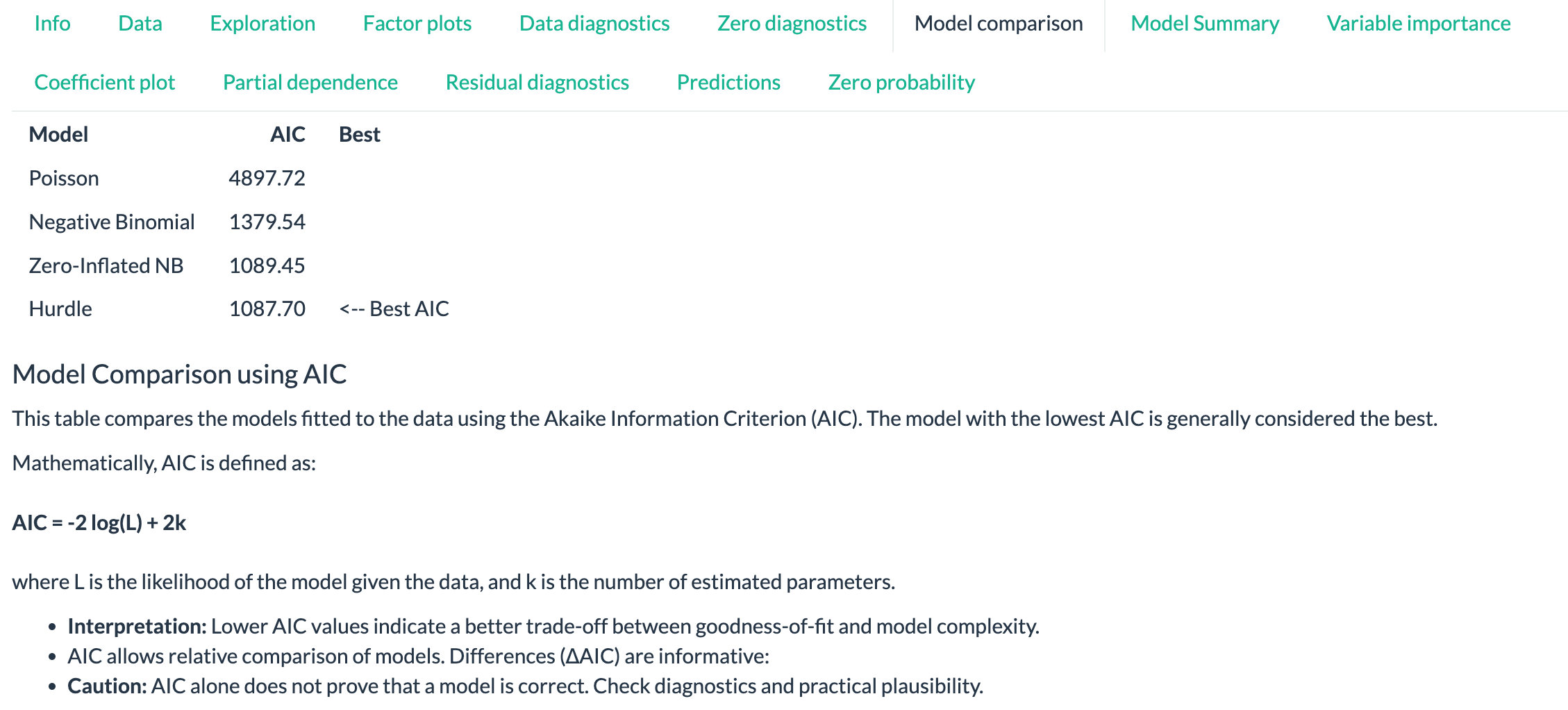}
    \caption{Model comparison tab shows every model AIC and highlight the model with lowest.}
    \label{fig:comparisonapp}
\end{figure}

After fitting the four candidate models, the application compares them using the Akaike Information Criterion (AIC),
\begin{equation}
\mathrm{AIC} = -2\log(L) + 2k,
\end{equation}
where $L$ is the maximised likelihood of the model and $k$ is the number of estimated parameters. The results are displayed in a comparison table containing the AIC value for each model. The model with the lowest AIC is highlighted as the best-supported candidate according to this criterion (Fig.~\ref{fig:comparisonapp}).

Although AIC provides a useful measure for relative model comparison, it should not be interpreted as definitive evidence that a model is correct. Instead, model selection should also consider residual diagnostics, parameter estimates, uncertainty, predictive performance, and ecological plausibility. This is especially important for ecological count data, where excess zeros, overdispersion, non-linear effects, and unbalanced categorical predictors may strongly influence model fit and interpretation.

\subsection{Model output, interpretation and diagnostics}

After the candidate models have been fitted and compared, the application provides a set of outputs to support model interpretation, diagnostic checking, and ecological inference. These outputs include the numerical model summary, variable importance plots, coefficient plots, partial dependence plots, residual diagnostics, observed-versus-predicted comparisons, and zero-part predictions for zero-inflated and hurdle models.

\subsubsection{Model summary}

\begin{figure}[htbp]
    \centering
    \includegraphics[width=0.8\textwidth]{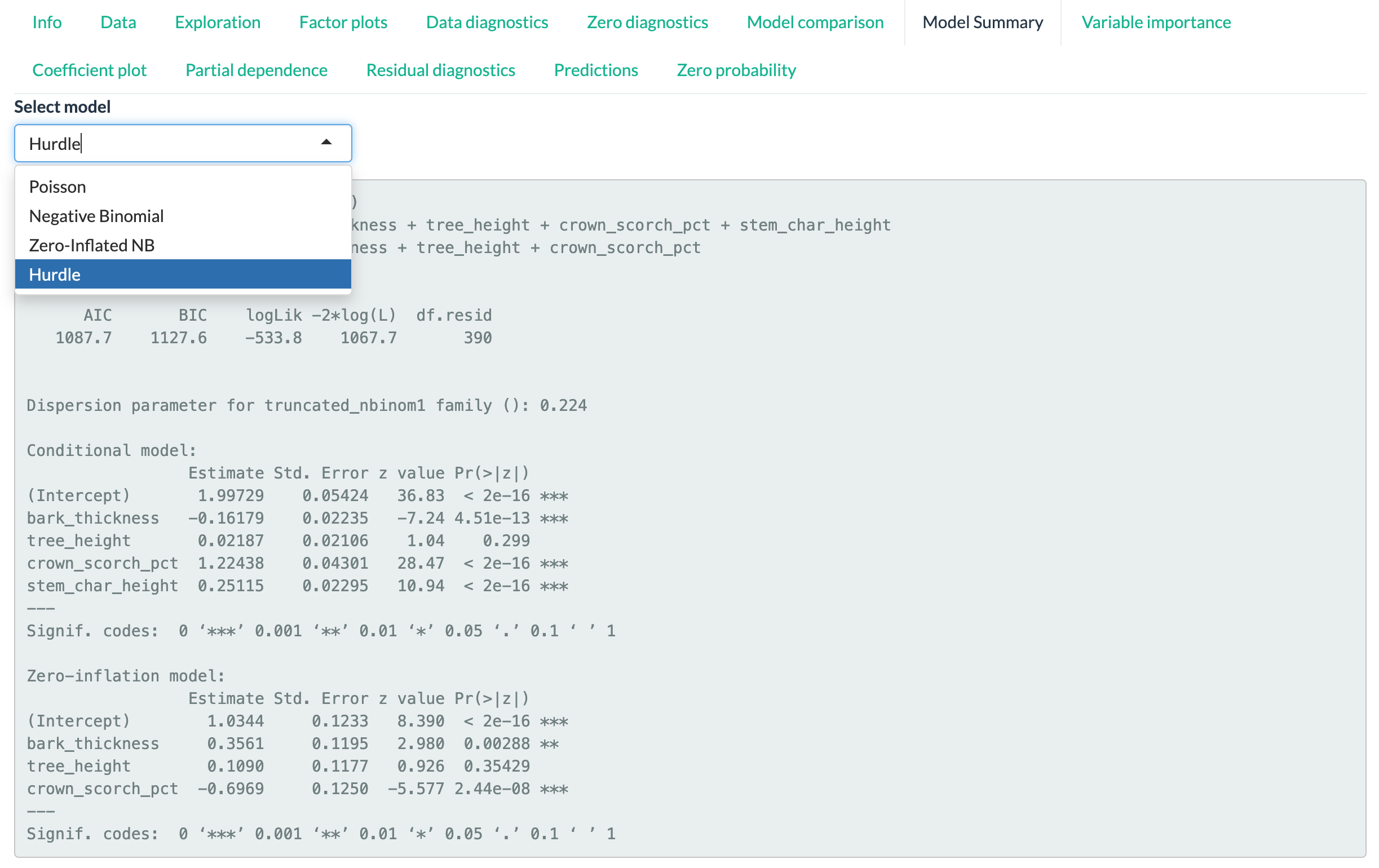}
    \caption{Model summary allow the user to select the models available and obtain the full model summary.}
    \label{fig:modelsummaryapp}
\end{figure}

For the selected model, the application displays the full model summary using the native \texttt{R} output (Fig.~\ref{fig:modelsummaryapp}). This summary provides the estimated regression coefficients, standard errors, test statistics, and associated significance values. For standard count models, such as the Poisson and negative binomial models, the coefficients describe the effect of each predictor on the expected response on the log scale. Positive coefficients indicate an increase in the expected count, whereas negative coefficients indicate a decrease.

For zero-inflated negative binomial and hurdle models, the summary contains two sets of coefficients. The first corresponds to the conditional count component, which explains variation in the magnitude of the response. The second corresponds to the zero component, which explains the probability of observing a zero outcome. Therefore, predictors may have different effects on the occurrence and intensity of the ecological process. For example, a variable may reduce the probability of observing a zero while simultaneously increasing the expected response among non-zero observations.

The interpretation of model coefficients is based not only on statistical significance, but also on effect size, uncertainty, and ecological plausibility. Since coefficients are estimated on the link-function scale, exponentiating them provides multiplicative effects on the expected response for the count component. This is particularly useful for interpreting the relative change in the response associated with a one-unit increase in a predictor, or with a one-standard-deviation increase when predictors have been standardised.

\subsubsection{Variable importance and coefficient plots}

\begin{figure}[htbp]
    \centering
    \includegraphics[width=0.8\textwidth]{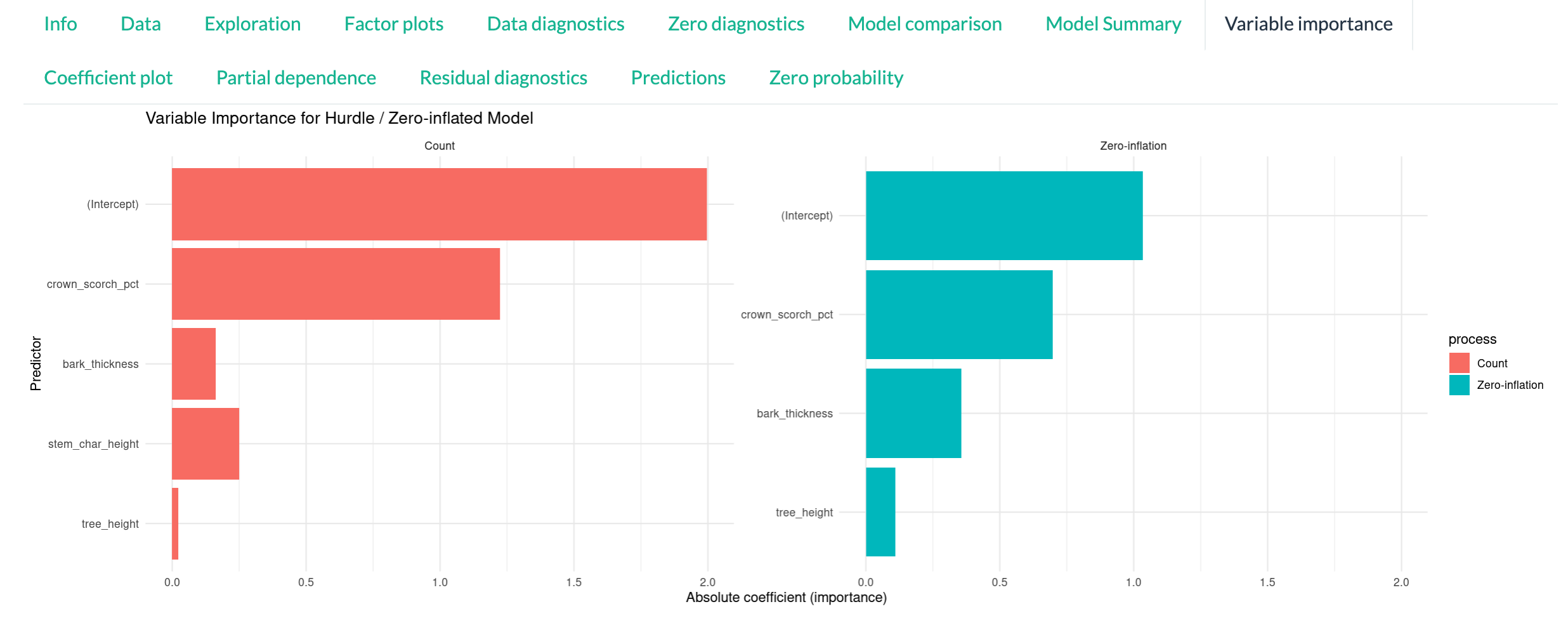}
    \caption{Variable importance tab showing variable importance for count and zero processes.}
    \label{fig:variableimpapp}
\end{figure}

To facilitate interpretation of fitted models, the application produces a variable importance plot based on the absolute magnitude of the estimated coefficients (Fig.~\ref{fig:variableimpapp}). For Poisson and negative binomial models, importance is calculated directly from the absolute values of the model coefficients. For zero-inflated and hurdle models, the plot is separated into the count and zero components, allowing users to distinguish predictors that influence the expected positive response from those that influence the probability of zero observations.

This measure of importance provides a simple visual summary of the relative strength of predictor effects. However, it should be interpreted with caution, as it reflects coefficient magnitude rather than statistical significance or predictive contribution. In particular, coefficient-based importance can be affected by the scale of the predictors; therefore, standardising numerical predictors can improve comparability across variables.

The coefficient plot provides a complementary visualisation by displaying the estimated coefficients together with approximate 95\% confidence intervals. For each coefficient, the interval is calculated as the estimate plus or minus 1.96 standard errors. A vertical reference line at zero is included to distinguish positive and negative effects. Coefficients with intervals that do not overlap zero provide stronger evidence of an effect, although interpretation should still consider ecological relevance and model assumptions.

\begin{figure}[htbp]
    \centering
    \includegraphics[width=0.8\textwidth]{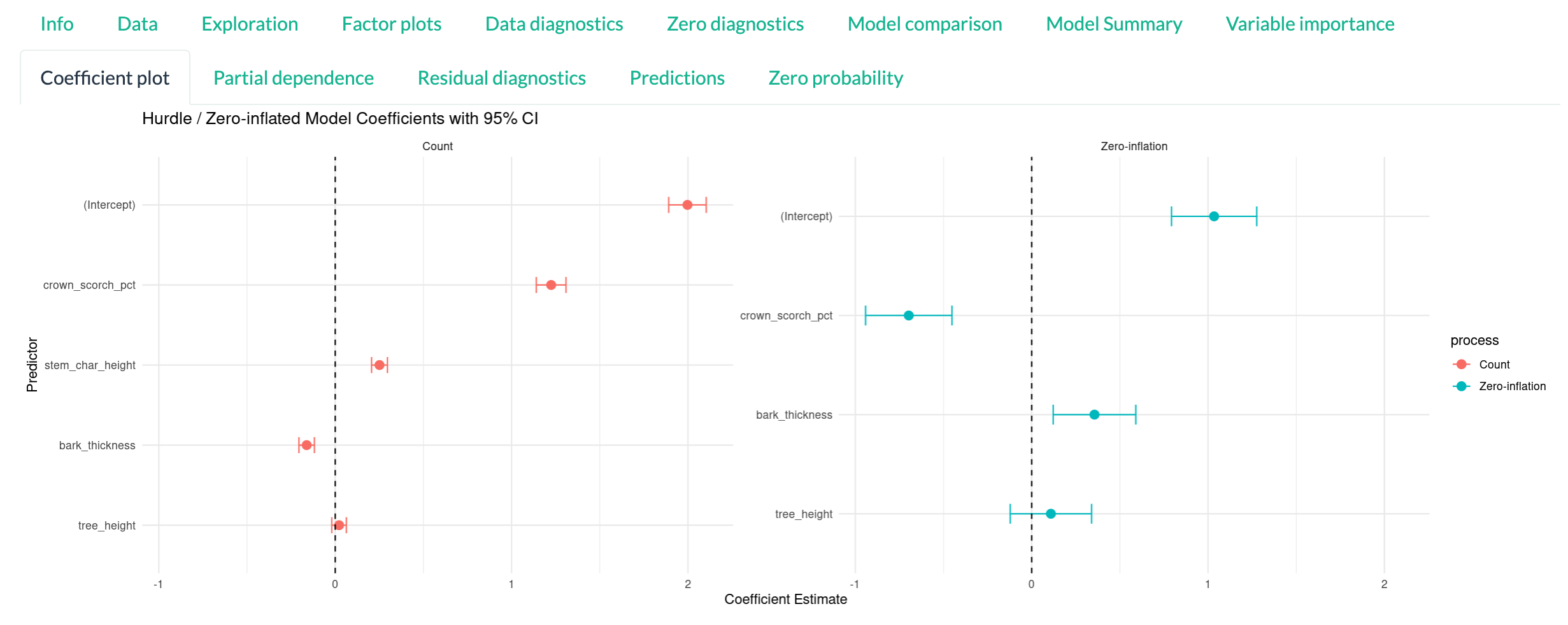}
    \caption{Coefficient plot tab showing estimated coefficients for count and zero processes.}
    \label{fig:coefplotapp}
\end{figure}

For zero-inflated and hurdle models, coefficient plots are faceted by model component (Fig.~\ref{fig:coefplotapp}). This allows the user to examine whether predictors have different roles in the count and zero-generating processes. In ecological applications, this distinction can be important because the mechanisms determining whether a response occurs may differ from those determining the magnitude of the response once it occurs.

\subsubsection{Partial dependence plots}

\begin{figure}[htbp]
    \centering
    \includegraphics[width=0.8\textwidth]{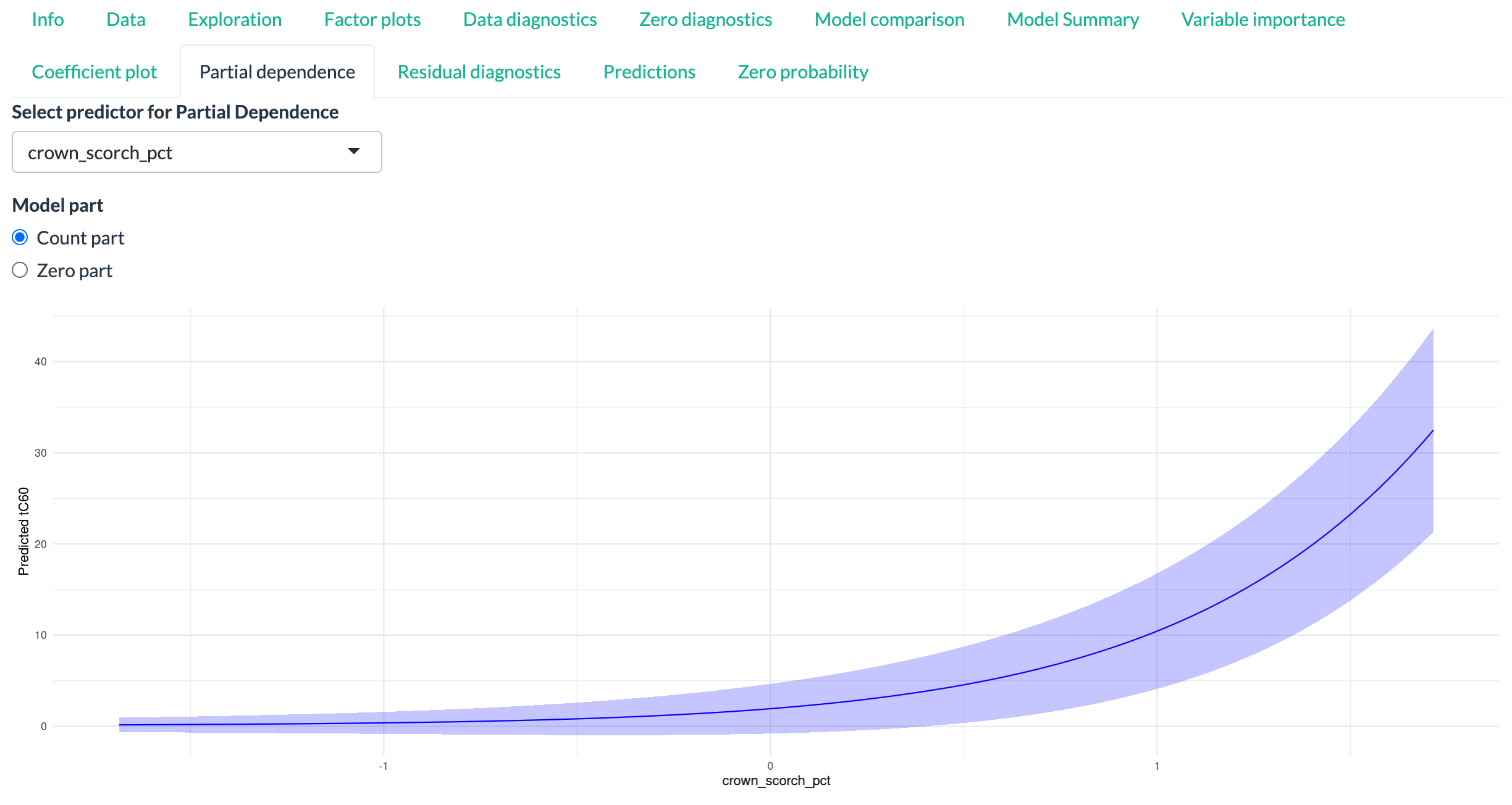}
    \caption{Partial dependence tab shows the marginal effect of individual predictors.}
    \label{fig:partialdependenceapp}
\end{figure}

The application also provides partial dependence (Fig.~\ref{fig:partialdependenceapp}) plots to visualise the marginal effect of individual predictors on model predictions. Users select a predictor of interest and, for zero-inflated or hurdle models, choose whether to display predictions for the count component or the zero component.

For numerical predictors, predictions are generated across a sequence of values spanning the observed range of the selected variable. Other numerical predictors are held at their mean values, while categorical predictors are fixed at a reference level. The resulting plot shows how the predicted response changes across the range of the selected predictor. For the count component, the plot displays the predicted response; for the zero component, it displays the predicted probability of a zero outcome.

\begin{figure}[htbp]
    \centering
    \includegraphics[width=0.8\textwidth]{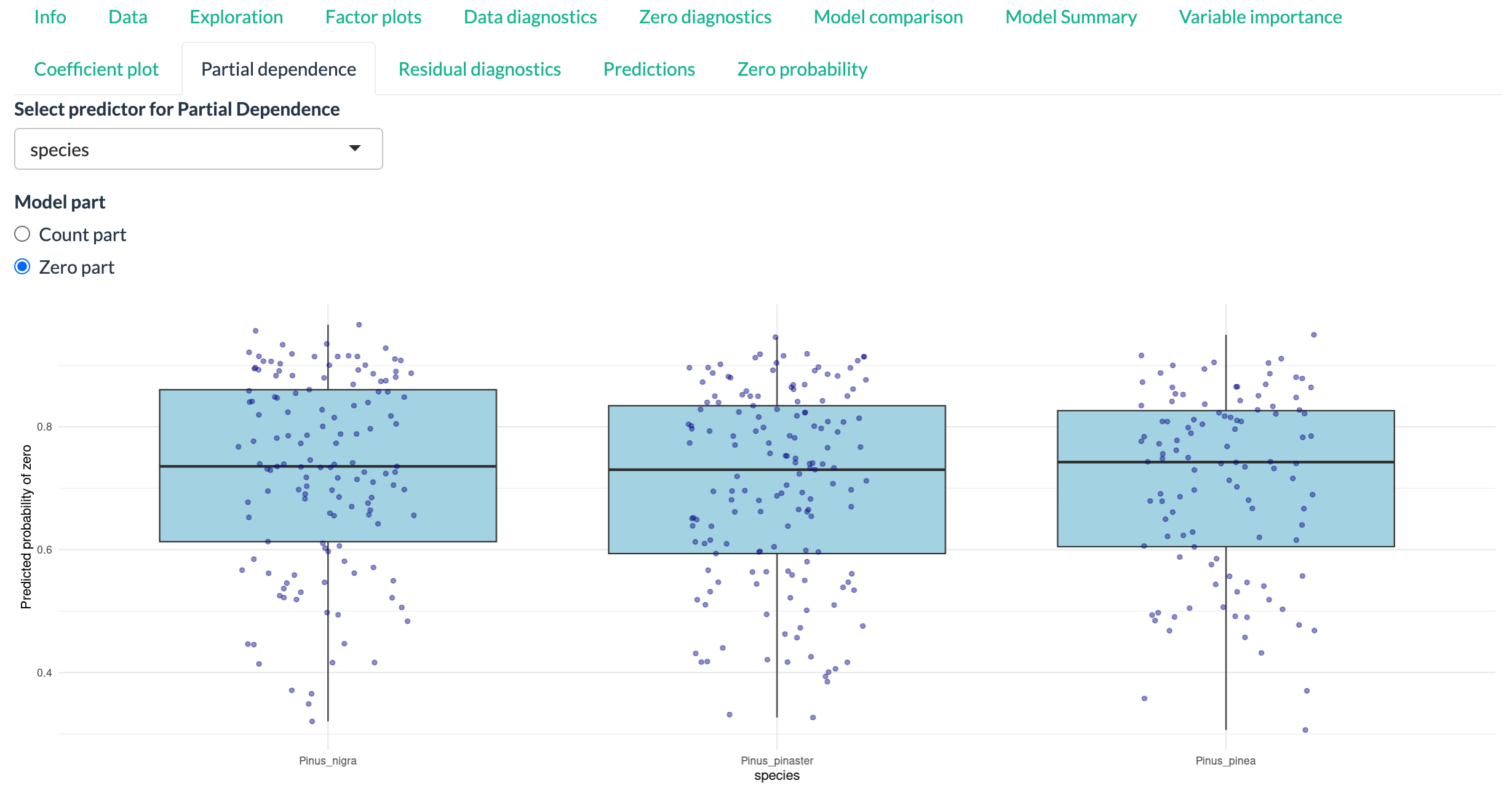}
    \caption{Partial dependence tab shows the marginal effect of individual predictors, in this case categorical variables.}
    \label{fig:partialdependencefactorapp}
\end{figure}

For categorical predictors, the application produces boxplots of predicted values across factor levels, with jittered points overlaid (Fig.~\ref{fig:partialdependencefactorapp}). This allows users to compare predicted responses among groups, such as species or burn season, while retaining information on the distribution of individual predictions.

Partial dependence plots are intended as interpretative tools that help translate model coefficients into response-scale patterns. They are particularly useful for identifying nonlinear-looking trends, differences among categorical groups, and predictors with strong effects. However, they should be interpreted in conjunction with coefficient uncertainty and model diagnostics.

\subsubsection{Residual and predictive diagnostics}

\begin{figure}[htbp]
    \centering
    \includegraphics[width=0.8\textwidth]{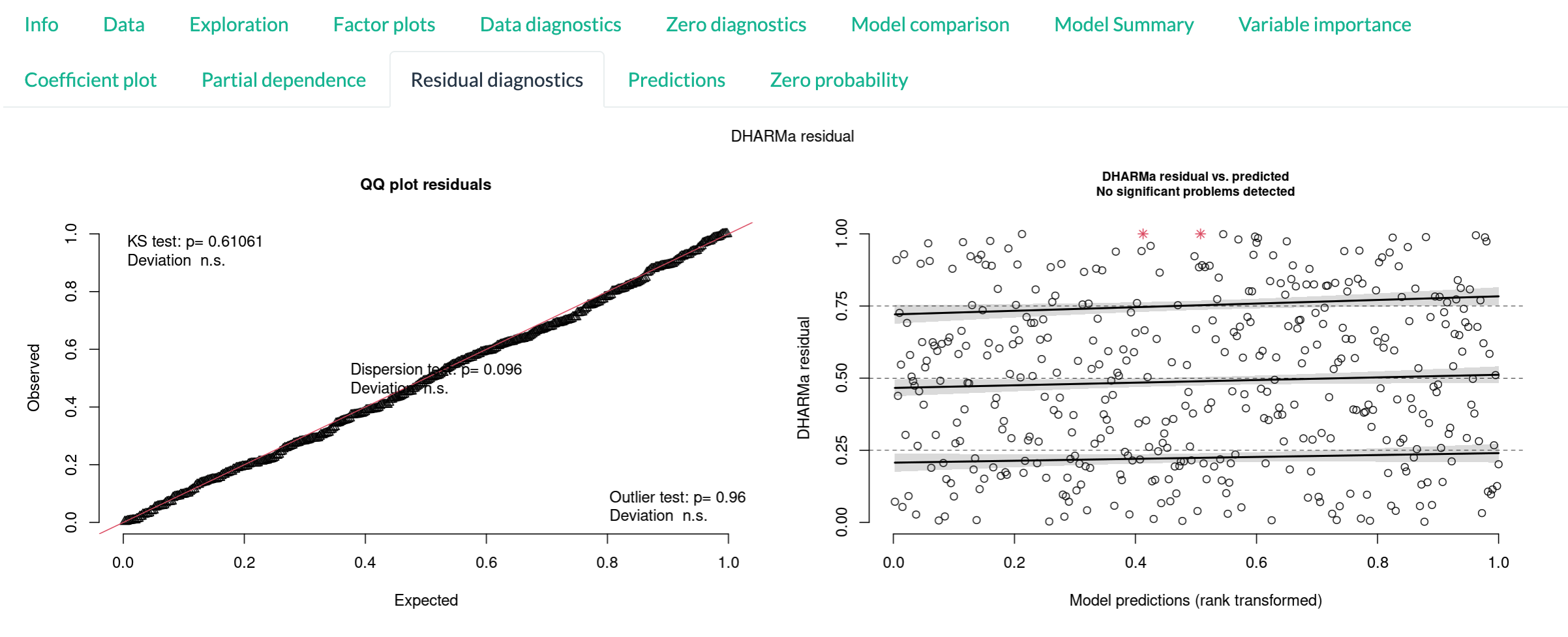}
    \caption{Residuals tab shows residual diagnostics generated with the \texttt{DHARMa} package.}
    \label{fig:residualsapp}
\end{figure}

Model adequacy is assessed using simulation-based residual diagnostics generated with the \texttt{DHARMa} package. These diagnostics provide a standardised framework for evaluating model fit across different count model classes. The diagnostic output includes a quantile--quantile plot, which assesses whether the simulated residuals follow the expected uniform distribution, and a residual-versus-predicted plot, which is used to detect systematic patterns in the residuals (Fig.~\ref{fig:residualsapp}).

A well-fitting model should show residuals that are approximately uniformly distributed and do not exhibit strong trends with the predicted values. Deviations from these expectations may indicate model misspecification, un-modelled heterogeneity, overdispersion, zero-inflation, influential observations, or missing covariates.

\begin{figure}[htbp]
    \centering
    \includegraphics[width=0.8\textwidth]{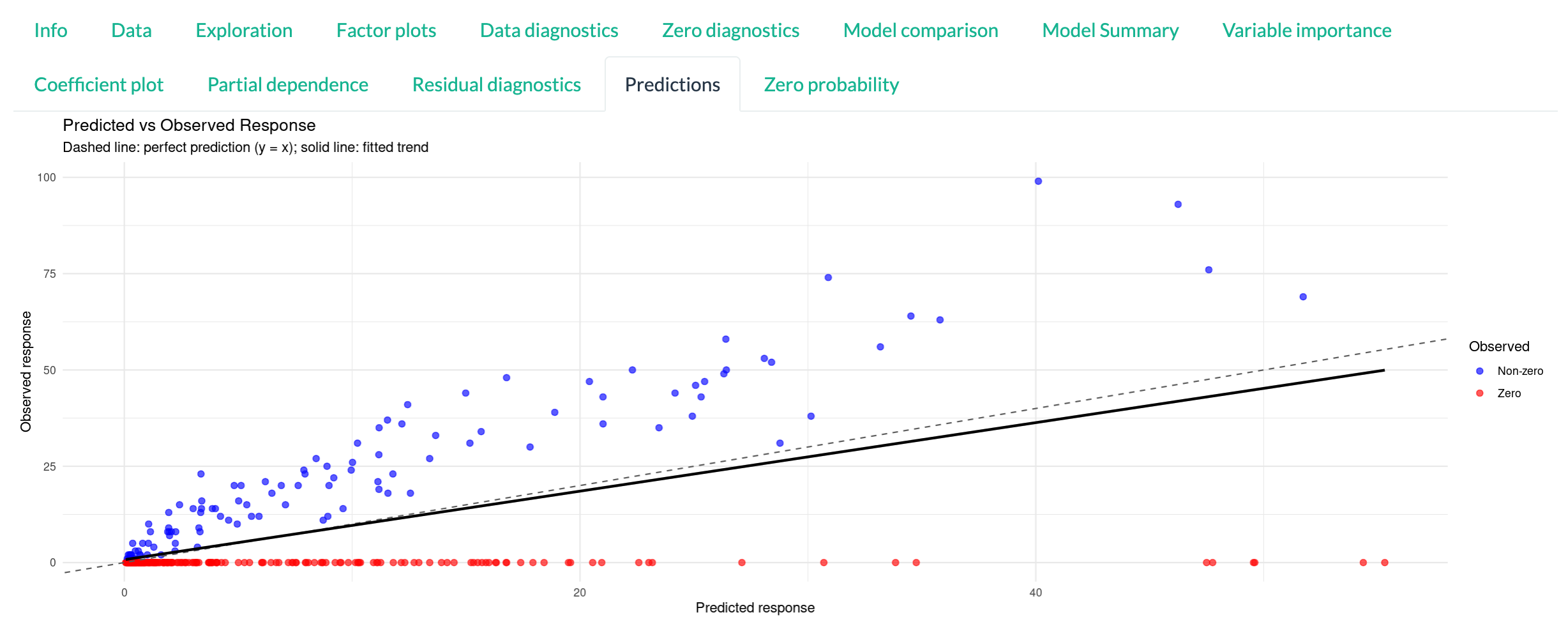}
    \caption{Predictions tab shows observed-versus-predicted.}
    \label{fig:predictionsapp}
\end{figure}

The application also displays an observed-versus-predicted plot (Fig.~\ref{fig:predictionsapp}). In this plot, each point represents one observation, with the predicted response shown on the horizontal axis and the observed response on the vertical axis. A dashed one-to-one line represents perfect agreement between predictions and observations, while a fitted trend line summarises the overall relationship between predicted and observed values. Points close to the one-to-one line indicate better predictive performance, whereas systematic deviations may indicate under-prediction or over-prediction in particular regions of the response distribution.

For ecological count data with many zeros, the plot distinguishes zero and non-zero observations. This is useful for assessing whether the selected model captures both the absence or non-occurrence process and the magnitude of positive counts.

\subsubsection{Zero-part predictions}

\begin{figure}[htbp]
    \centering
    \includegraphics[width=0.8\textwidth]{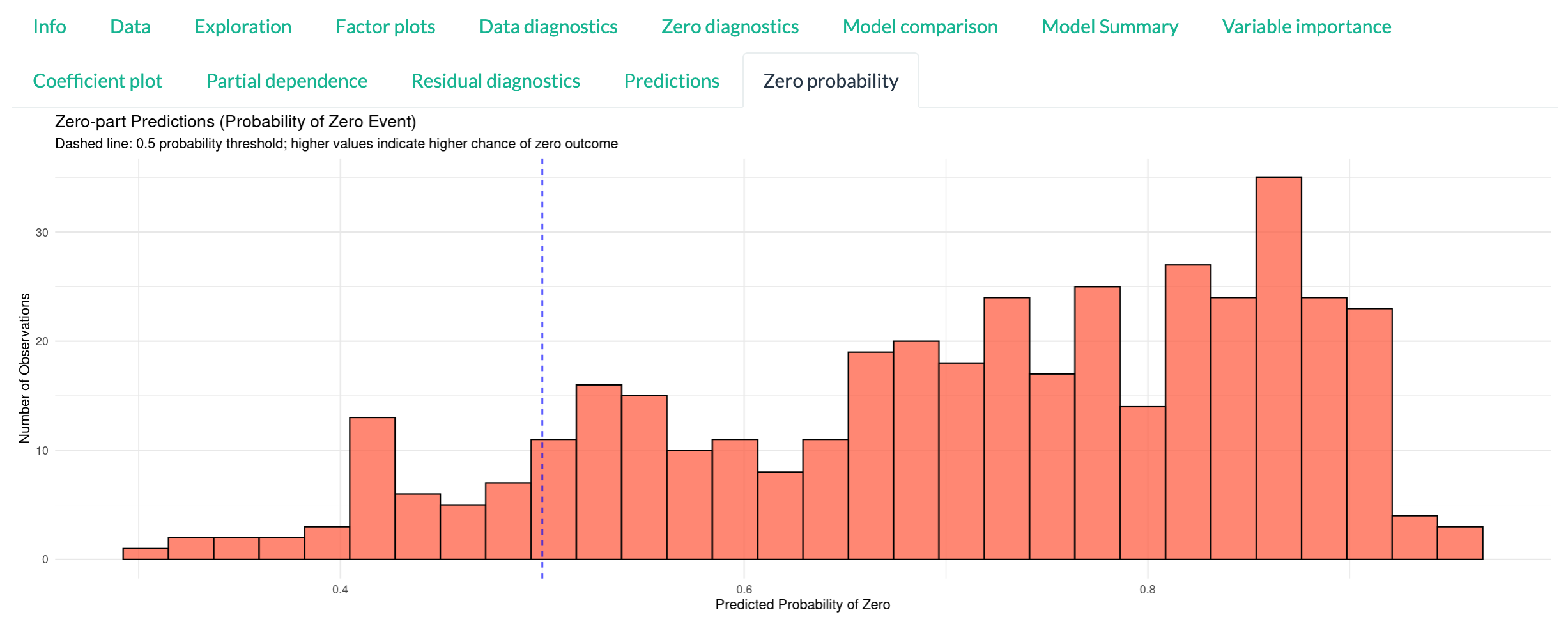}
    \caption{Zero probability tab shows the predicted probability of a zero outcome.}
    \label{fig:zeroprobabilityapp}
\end{figure}

For zero-inflated and hurdle models, the application provides an additional diagnostic plot showing the predicted probability of a zero outcome (Fig.~\ref{fig:zeroprobabilityapp}). This plot is generated only for models with an explicit zero component. The distribution of predicted zero probabilities is displayed as a histogram, with a reference line at 0.5 to indicate observations that are more likely than not to be predicted as zeros.

Predicted probabilities close to zero indicate observations for which the model expects the ecological event or response to occur, whereas probabilities close to one indicate observations with a high probability of a zero response. The shape of this distribution provides information on how strongly the model separates zero and non-zero observations. For example, a concentration of predicted probabilities near 0.5 may suggest weak separation, while a broader distribution with values closer to 0 or 1 may indicate stronger discrimination between zero and non-zero outcomes.

More generally, zero-part predictions provide insight into the conditions associated with a low probability of response occurrence. By identifying observations, sites, groups, or environmental conditions with a high predicted probability of zero outcomes, these outputs help clarify the ecological processes that may prevent or limit the response. They therefore support both model interpretation and evidence-based ecological decision-making.

\section{Conclusions}

This paper presented an interactive modelling workflow for the analysis of count data, with particular emphasis on responses characterised by overdispersion and excess zeros. The application was designed to guide users through the main stages of a reproducible statistical analysis, including data upload, exploratory visualisation, descriptive diagnostics, model fitting, model comparison, coefficient interpretation, prediction, and residual checking. By integrating these steps within a single interface, the tool provides a transparent and accessible framework for analysing ecological responses that may arise from multiple underlying processes.

A key feature of the application is the ability to compare standard and extended count models, including Poisson, negative binomial, zero-inflated negative binomial, and hurdle formulations. This is particularly relevant in ecological applications, where zero observations may reflect true absence, lack of exposure, non-detection, or other structural processes \citep{martin2005zero, warton2005many}. The inclusion of both zero-inflated and hurdle models allows users to explore alternative hypotheses about the mechanisms generating zeros and positive counts, rather than treating all observations as arising from a single homogeneous count process.

The workflow also emphasises model interpretation and diagnostic evaluation. Model summaries, coefficient plots, variable importance measures, partial dependence plots, predicted-versus-observed comparisons, and zero-probability plots are provided to help users interpret fitted models in both statistical and ecological terms. In addition, simulation-based residual diagnostics are included to assess model adequacy and identify potential departures from model assumptions \citep{hartig2024dharma}. These outputs encourage users to move beyond model selection based only on information criteria and to evaluate whether fitted models are both statistically adequate and ecologically plausible.

The proposed application is therefore intended to support research by making advanced count-data modelling tools more accessible to researchers and practitioners. It provides a flexible environment for exploring complex datasets, identifying potential data issues, comparing competing model structures, and communicating model results through interpretable visual outputs. Although the example presented here uses a simulated post-fire tree damage dataset, the same workflow can be applied to a wide range of count responses, including abundance, event counts, damage scores, incidence data, or exposure durations.

Future developments could extend the application by incorporating mixed-effects structures, spatial or temporal random effects, Bayesian model fitting, automated cross-validation, and additional tools for uncertainty propagation. These extensions would be particularly valuable for ecological datasets with hierarchical sampling designs, spatial dependence, temporal replication, or strong site-level heterogeneity. Nevertheless, the current implementation provides a practical and reproducible foundation for analysing zero-heavy count data and for supporting statistically informed decision-making.

\bibliographystyle{apalike}
\bibliography{cas-refs}

\end{document}